# Three-dimensional micro-electromagnet traps for neutral and charged particles

M. Drndic, C.S. Lee, and R.M. Westervelt

*Department of Physics and Division of Engineering and Applied Sciences*

*Harvard University, Cambridge, Massachusetts, MA 02138*

## Abstract

Three-dimensional (3D) micro-electromagnets were developed to control particle motion in magnetic field landscapes in vacuum near a chip. Multiple layers of micron-scale conductors separated by transparent insulators create particle containers with deep, symmetric and time-dependent potentials, suitable for integration in quantum circuits. Single and coupled multiple 3D traps, integrated 3D traps with a mirror, 3D guides for neutral particles with spin, and traps for electrons in vacuum are described.

With advances in particle cooling techniques [1] and microfabrication of complex structures [2], it is now feasible to study the quantum transport of low-energy atoms, molecules, and electrons inside low-dimensional structures of size approaching their wavelength. Quantum transport of electrons inside semiconductor and metal nanostructures has shown what is possible [2]. The control of particles in vacuum can be achieved on a microscopic scale using electromagnetic field patterns, opening possibilities ranging from integrated matter-wave optics and quantum dots in vacuum to quantum computation. Manipulation schemes for atoms above planar microstructures have been suggested [3]. Those realized experimentally are planar micro-electromagnet atom mirrors [4], micro-traps [5], and atom guides [6,7]. Quantum wires and quantum dots for atoms have been proposed using planar structures [8]. Conductance quantization of atom flow through a constriction has been predicted [9], analogous to electron flow through quantum point contacts [10]. Microtraps for ions modeled on Paul traps have been realized [11], and two-ion super- and subradiance has been observed [12]. Manipulation of electrons in vacuum for energies $E > 1$ eV has been realized for micro-vacuum tubes and flat panel displays [13]. Electrons with lower energy will permit study of low-dimensional gases in the ballistic regime in vacuum.

In this Letter we describe the design and fabrication of three-dimensional (3D) micro-electromagnets (µems) for the control of neutral particles with spin (atoms and molecules). Multiple layers of metal conductors separated by transparent insulators create micron-scale containers: (i) 3D µem particle traps and guides, (ii) integrated 3D µem traps with particle mirrors, and (iii) coupled multiple 3D µem traps. The feasibility of µem traps for electrons in vacuum is demonstrated. Three dimensional µems create deep and spatially symmetric potentials that can be integrated into circuits for quantum transport and quantum computing. Using modern fabrication techniques, µems can be precisely designed and constructed. Possible applications extend to nuclear magnetic resonance and biological systems.

Three dimensional µem traps and channels can create tightly confining quantum systems. The wavelength of cold atoms is $\lambda_{dB} = 3.1/\sqrt{AT}\,(\mu m\sqrt{\mu K})$ with atomic number $A$ and temperature $T$. For cesium $\lambda_{dB} = 0.1 \mu m$ at $T = 10$ μK, achievable in a magneto-optical trap. In planar microstructures, particle confinement is typically weakest perpendicular to the plane [14]. Three dimensional μem configurations can offer stronger confinement potentials and higher curvatures. Consequently, the condition for quantum behavior $\Delta E > k_B T$, with $\Delta E$ the energy level separation and $k_B T$ the thermal energy, is easier to realize. Three dimensional structures can also create spatially symmetric trapping potentials.

Figure 1 shows schematic diagrams, magnetic field profiles, and micrographs for 3D μems for neutral atoms. Figure 1(a) shows a quadrupole microtrap consisting of two stacked current loops with a circular hole through the trap center. This is a miniature anti-Helmholtz trap [15]. Figure 1(b) shows a 3D atom guide consisting of two stacked pairs of parallel conductors. Figure 1(c) presents a 3D trap with an additional pair of rings to create a bias magnetic field, placed on top of a μem mirror [4], which can be used to load the trap as described below. Magnetic field profiles which determine the particle potential are shown in the middle column of Fig. 1. The potential energy is $U = g_F m_F \mu_B B$, with magnetic field magnitude $B$, Bohr magneton $\mu_B$, Landé g factor $g_F$, and magnetic quantum number $m_F$, which remains constant for the adiabatic case where particle motion is slow enough to prevent spin flip. Weak-field seeking particles ($m_F g_F > 0$) are attracted to $B$-field minima.

Micrographs of fabricated 3D μems are shown in the right column of Fig. 1. Each structure consists of layers of metal conductors separated by transparent insulators, with cavities to permit atom motion. The devices were fabricated on sapphire plates using multiple steps of thin film deposition and optical lithography, which can produce features as small as 150 nm. The transparent insulator films were spun-on polyimide [16] (magnetic permeability $\mu_r = 1$ and dielectric constant $\varepsilon_r = 3.3$) and the metal films were evaporated gold. The structures allow optical access to atoms through the transparent substrate and insulator layers, permitting detection, imaging and laser manipulation of atoms and molecules. The fabrication method can integrate devices both horizontally along the chip and vertically to make circuits for particle motion.  These circuits avoid atom heating by lasers, found in optical manipulation of atoms [15].

Magnetic field patterns, illustrated in Fig. 1, are produced in 3D µems by electric currents.  The current density obtained in structures like those in Fig. 1 at room temperature can reach ~$10^8$ A/cm$^2$; cooling to liquid nitrogen or helium temperatures can reduce wire resistance and increase current density.  Figure 1 (a) shows magnetic field contours for a 3D trap with two conductor loops of radius $R$ a distance $d$ apart.  The trap depth $B_{max} = 0.5$ $I/R$ (Tµm/A) is greatest for $d =1.25R$, corresponding to maximum particle temperature $T_{max}= \mu_B B_{max}/k_B = 0.3$ $I/R$ (Kµm/A).  The field gradient and curvature scale as $I/R^2$ and $I/R^3$ respectively.  Figure 1 (b) shows field contours for a guide with four conductors a distance $d$ apart.  The guide depth is $B_{max} = 0.64$ $I/d$ (Tµm/A) corresponding to $T_{max}= 0.4$ $I/d$ (Kµm/A) for $\mu = \mu_B$.  Stronger traps and guides are formed by 3D µems than by 2D µems [7] of comparable size and power dissipation.  Table I compares 3D and 2D traps and guides for cesium atoms, showing trap depth $T_{max}$, field gradients $\partial B/\partial z$ and $\partial B/\partial r$, and energy level spacings $\hbar\omega_z$ and $\hbar\omega_r$.  Three dimensional µems are favored for superconducting conductors, because the critical current density ~ $10^7$ A/cm$^2$ [17] and magnetic field are lower than achievable in gold.

The magnetic field pattern from µems can be changed very rapidly, creating new possibilities for particle manipulation.  For the 3D trap in Fig. 1(a) the time constant is $L/R = 20$ ps, with inductance $L = 0.1$ nH and resistance $R = 5$ Ω.  The electrical control of magnetic fields is an advantage over permanent-magnet structures, which require physical motion.  Figure 1 (c) illustrates a time dependent µem structure consisting of a 3D µem trap, with an additional pair of coils to produce a bias magnetic field, placed on top of a µem mirror [4].  By modulating the mirror current, this structure can cool atoms and load them into the trap.  Adiabatic compression of atoms into small volumes increases their temperature because the magnetic potential is conservative.  However, a receding mirror can cool atoms before loading into a trap.  A ball bouncing elastically from a hard wall moving away will stop when the wall velocity is $v/2$, where $v$ is the incoming ball velocity.  For a µem mirror this velocity is achieved by decreasing the mirror current exponentially $I(t) = I_o e^{-\pi vt/a}$, where $a/2$ is the distance between mirror wires. The time required to cool is ~ 70 μs for a mirror with $a$ = 200 μm [4] and initial atom velocity $v$ = 1 m/s for atoms dropped from a larger magneto-optical trap above. The mirror current can be modulated in other ways at high speed to manipulate trapped atoms, including periodic kicks to study the nonlinear dynamics of atom motion [18].

Figure 2 shows a 3D μem which forms a double trap with adjustable trap coupling as well as a toroidal trap. Both structures are interesting for quantum computing [21,23]. As shown in Fig. 2(a) the trap is composed of three coaxial current loops of radius $R$ separated by equal distances $d = R$. The evolution of trap geometry is shown by plots of magnetic field magnitude in Figs. 2(b) to 2(f) as the ratio $I_1/I_2$ is increased from 0.8 to 2.2, with $I_1$ the current in the top and bottom loops and $I_2$ the current in the middle loop. Particles are trapped in low magnetic field regions shown in red. Figures 2 (b) to (d) show two traps coupled by adjustable inter-trap tunneling controlled by $I_1/I_2$, Fig. 2 (e) shows a single simply connected trap, and Fig. 2 (f) shows a single toroidal trap of variable aspect ratio.

Adjustable double traps are a building block of quantum circuits for atoms. For the sequence of double traps Figs. 2 (b) to 2(d), the individual trap depth is largest for $I_1/I_2 = 0.73$ ($B_{max} \approx 30$ mT for $d = R = 1$ μm and $I_2 = 0.1$ A). The potential barrier is lowered in Fig. 2(d) where the traps are pushed together with increased inter-trap tunneling, forming states analogous to molecular bonds. Artificial molecules have been formed in tunnel-coupled semiconductor quantum dots with electrons [19]. In the case of atoms, the formation of such bonds could be characterized spectroscopically. As in the $NH_3$ molecule, the lowest energy states are symmetric and anti-symmetric superpositions of the two single trap wavefunctions [20]. By adding more loops one can construct an array of coupled traps where particles can move from one trap to the next. These could be a new tool to entangle neutral atoms for quantum computation [21].

Once the double traps join, a single trap is formed as shown in Fig. 2 (e). Here the field magnitude $B$ increases quadratically away from the center, unlike the linear increase for single two-coil traps in Figs. 1(a) and 2(b). The trap depth for Fig. 2(e) is $B_{max} \approx 80$ mT for $I_1/I_2 = 1.4$, $d = R = 1$ µm, and $I_2 = 0.1$ A, and $B = 0$ at the trap center. For higher ratios, $I_1/I_2 > 1.4$, $B$ becomes nonzero at the center. A spherical hexapole three-coil trap with $B = 0$ in the trap center [22] is obtained if the radius of the outer rings and the separation $d$ are both $R/\sqrt{2}$, and $I_1/I_2 = 1$.

A toroidal trap occurs when the current ratio is $I_1/I_2 > 1.8$, as shown in Fig. 2 (f). Toroidal traps are a possible approach for quantum computers based on coupled ions [23]. Micro-electromagnet toroidal traps could confine cold plasma into small volumes $\sim 10^{-9}$ mm$^3$. Recently, cold neutral plasma consisting of electrons at 100 mK and ions at 10 µK has been produced [24]. For Fig. 2(f) the trap depth is $B_{max} \approx 16$ mT, with $R = 1$ µm, $I_1 = 0.1$ A and $I_1/I_2 = 2.2$. As the current ratio $I_1/I_2$ increases above 2.2, the trap is pushed further out and the trap depth and gradient increase.

Three-dimensional µems can be used for other applications. Coils like those in Fig. 2 (a) could be used to conduct nuclear magnetic resonance experiments [25] with large field gradients $\sim 10^4$ T/m. Arrays of 3D µems could arrange small biological organisms whose population and characteristics depend on local magnetic field. Three dimensional µems with voltage electrodes could be used for trapping and guiding of ions, electrons and polar molecules.

Finally, µems can trap electrons in vacuum, forming quantum dots or artificial atoms interesting for new experiments. Vacuum traps minimize interference from outside effects found for semiconductor quantum dots, and could create more perfect structures for the study of low-dimensional systems with possible applications in vacuum integrated circuits and approaches to quantum computers [26]. Electron motion in traps is 2 to 3 orders of magnitude faster than atom traps due to their lower mass, an advantage for applications.

Figure 3 illustrates a simple electron microtrap composed of a conducting ring of radius $a$ at voltage $V_o > 0$ located in a homogeneous magnetic field $B$ along the $z$–axis. Near $z = 0$, this approximates a Penning trap [27], consisting of a quadrupole electrostatic field binding the electron in the $z$- direction and a magnetic field creating a circular orbit about the $z$-axis. The electron is attracted to the positive ring and trapped when the magnetic field magnitude $B > B_{min} \approx (2mV_o/e)^{1/2}/a$, with $m$ and $e$ the electron mass and charge. Inside the trap, the electron motion can be decoupled into three independent harmonic oscillators [27]: axial, cyclotron, and magnetron, with frequencies $\omega_z$, $\omega_{c'}$, $\omega_m$ and lifetimes $\Gamma_z$, $\Gamma_{c'}$, $\Gamma_m$, respectively. Due to the dissipative magnetron motion, the trap is metastable and the electron will drift out after the magnetron lifetime $\Gamma_m$. For $a = 1\mu m$, $V_o = 0.1$ V and $B = 4$ T, $\omega_z/2\pi = 20$ GHz, $\omega_{c'}/2\pi = 110$ GHz, $\omega_m/2\pi = 2$ GHz, $\Gamma_z \approx 2$ s, $\Gamma_{c'} \approx 0.2$s, and $\Gamma_m \approx 5$hr. The cyclotron and axial radiation can be in thermal equilibrium with the background radiation field. Electrons are available from a variety of sources: photo, thermal, field emission or even radioactive sources. Detection and measurement techniques could span from capacitance and charge measurement as in semiconductor quantum dots [2] to resonant frequency techniques as in macroscopic Penning traps [27].

We thank C.H. Tseng, G. Gabrielse, and M. Prentiss for helpful discussions. This work was supported by NSF grant DMR-9809363, NSF grant PHY-9871810, and ONR Grant N00014-99-1-0347.

**Figure and Table Captions**

Table I - Comparison of 3D and 2D µem traps and guides for cesium atoms showing trap depth $T_{max}$, field gradients $\partial B/\partial z$ and $\partial B/\partial r$, and energy level spacings $\hbar\omega_z$ and $\hbar\omega_r$. For 2D traps, two coaxial loops of radii $R_1$ and $R_2$, carry currents $I_1$ and $I_2$, optimized for maximum depth ($I_2/I_1$ = 1.87 and $R_2/R_1$ = 2.66) [14]. For 2D guides [7], four parallel wires separated by distance $d$ carry currents $I_1$ in the inner pair and $I_2$ in the outer pair, optimized for maximum depth ($I_2/I_1$ = 2.1). The power dissipation and size of 3D and 2D cases are the same ($R = R_1$). The field gradients are calculated for dimensions $R = d = 1$ µm, current $I = 0.1$ A and magnetic moment $\mu = \mu_B$. A bias field $B_{bias} = 1 mT$ in the $z$ direction ensures adiabaticity.

FIG. 1 (color) - Three-dimensional micro-electromagnet (µem) schematic diagrams, contours of constant magnetic field magnitude $B$ about the trap or guide center, and device photographs for (a) trap with two rings of radii $R$ and separation $d$, with opposite currents $I$, (b) guide with four conductors, (c) trap with bias field coils above a µem mirror, $d_1 = 0.25R$, $d_2 = 1.5R$; upper contour plot for trap off, mirror on, lower contour plot for trap on, mirror off. Magnetic field contours are spaced by (a) $\Delta B = 0.03 I / R$ (Tµm/A); (b) $\Delta B = 0.05 I / R$ (Tµm/A) and (c) (above) $B_{n+1}/B_n = 0.8$ where n = 1 to 20, (below) by $\Delta B = 0.03 I / R$ (Tµm/A).

FIG. 2 (color) - Three-dimensional micro-electromagnet trap with three current loops: (a) schematic diagram, (b) to (f) contours of constant magnetic field magnitude $B$ about the trap center. (b) to (d) Two coupled traps are formed with a variable barrier as the ratio of currents in the two outer rings $I_1$ and the middle ring $I_2$ is varied: (b) $I_1/I_2 = 0.8$, (c) $I_1/I_2 = 1.0$, and (d) $I_1/I_2 = 1.2$. (e) A single trap is formed for $I_1/I_2 = 1.4$. (f) A toroidal trap is formed for $I_1/I_2 = 2.2$. Magnetic field contours are spaced by $\Delta B = 0.1$ T for $R = d = 1$ µm.

FIG. 3 - Contours of constant electric potential about the center of a micro-electromagnet electron trap consisting of a conductor ring with radius *a* held at voltage $V_o$.

Table I

|  | $T_{max}$ (mK) | $\partial B/\partial z$ ($10^4$ T/m) | $\hbar\omega_z$ (10 μK) | $\partial B/\partial r$ ($10^4$ T/m) | $\hbar\omega_r$ (10 μK) |
|---|---|---|---|---|---|
| 3D trap | 30 | 10 | 16 | 5 | 8 |
| 2D trap | 4.5 | 3 | 5 | 1 | 1.6 |
| 3D guide | 40 | 15 | 23 | 15 | 23 |
| 2D guide | 8 | 4 | 6 | 4 | 6 |

Figure 1

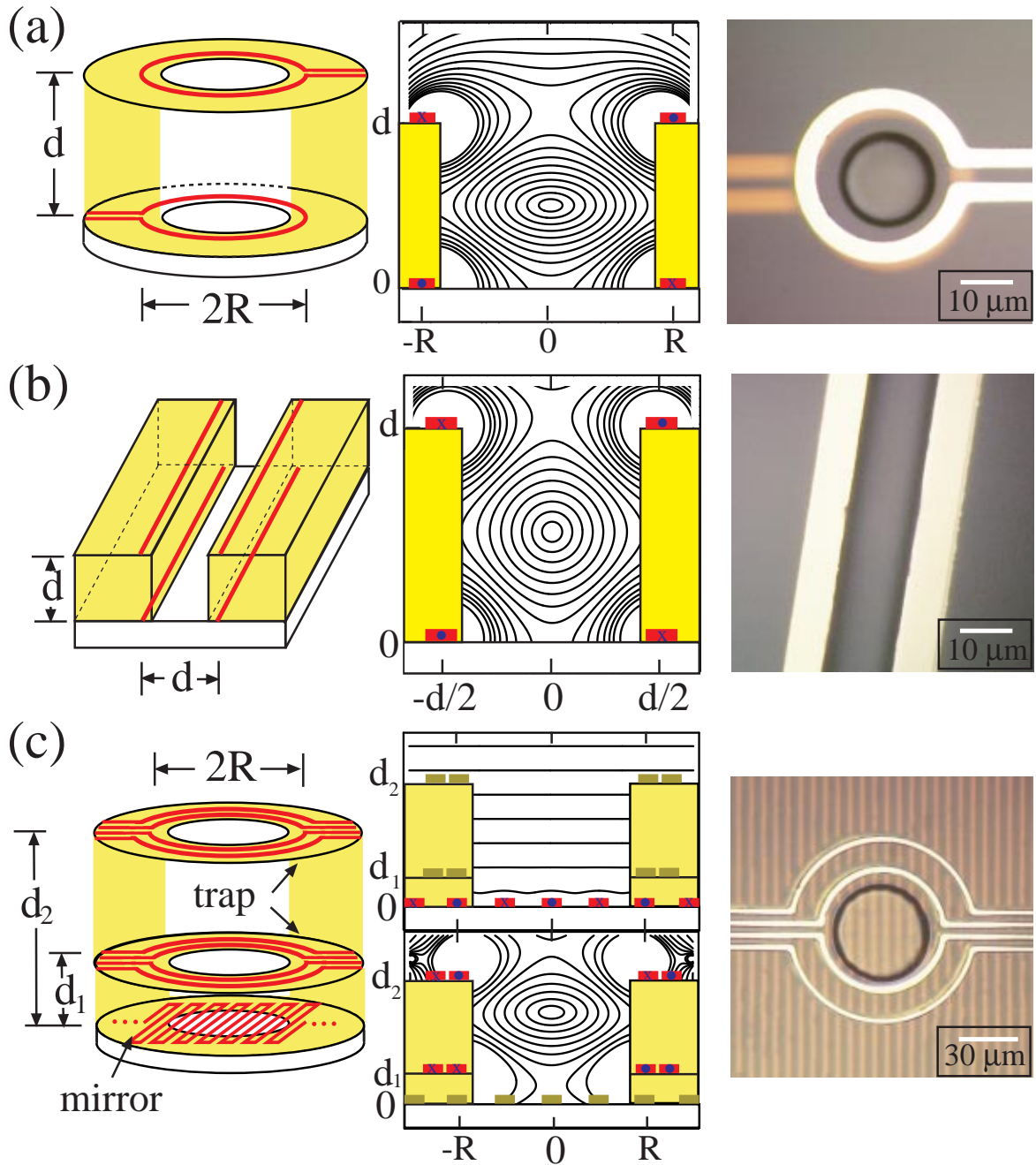

Figure 2

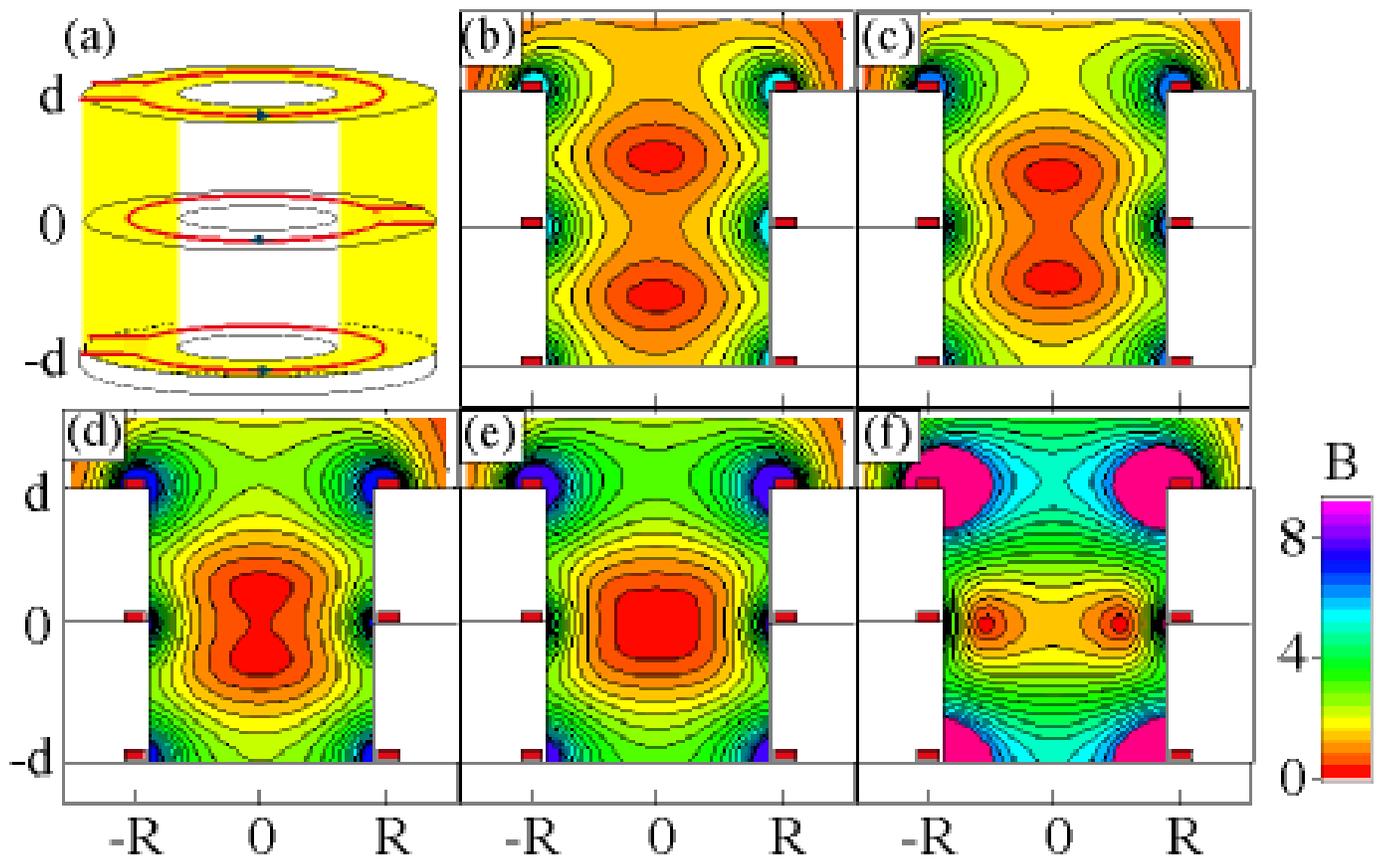

Figure 3

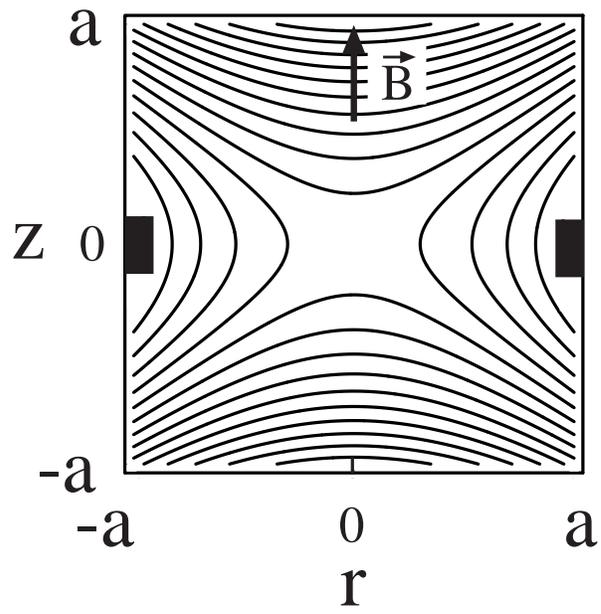